\documentclass[11pt]{article}
\usepackage{geometry} % See geometry.pdf to learn the layout options. There are lots.
\usepackage{apacite}
\usepackage{lscape}  % for landscape rotation of large figures
\usepackage{graphicx} % for graphs 
\usepackage{topcapt} % for tables (caption at top)
\usepackage{amssymb}
\usepackage{amsmath}
\usepackage{wrapfig} % to include figure with text wrapping around it
\usepackage[margin=10pt,font=small,labelfont=bf]{caption} % for improved layout of figure captions with extra margin, smaller font than text
\usepackage{fancyhdr} % for better header layout
\usepackage{eucal}
\usepackage[english]{babel}
\usepackage[usenames, dvipsnames]{color}
\usepackage[perpage, symbol, bottom]{footmisc}
\usepackage[round, sort, numbers, authoryear, sectionbib]{natbib}
\usepackage{ifthen}
\usepackage{multicol} % for pages with multiple text columns, e.g. References
\usepackage[nottoc]{tocbibind} % correct page numbers for bib in TOC, nottoc suppresses an entry for TOC itself
\usepackage{natbib}
\usepackage{verbatim}
\usepackage[compatible]{nomencl}
\usepackage{float}  % allows changing of default float placements
\usepackage{flafter} % prevents floats appearing before their reference
\usepackage{subcaption}
\usepackage{underscore}
\usepackage{titlesec}
\usepackage{lscape}
\usepackage{authblk}
\usepackage{etoolbox}
\usepackage[sectionbib]{chapterbib}
\usepackage{pdfpages}
\usepackage{tabulary}
\usepackage{longtable}
\usepackage{array}
\usepackage{booktabs}
\usepackage{lscape}
\usepackage{soul}
\usepackage{siunitx}
\usepackage{bigints}
\usepackage{pifont}
\usepackage{tabu}
\usepackage{bbm}
\usepackage{color, colortbl}
\usepackage[table, svgnames, dvipsnames]{xcolor}
\usepackage{fancyhdr}

\newcommand{\ra}[1]{\renewcommand{\arraystretch}{#1}}
\renewcommand\thepage{\arabic{page}}
\newcommand*\colourcheck[1]{%
  \expandafter\newcommand\csname #1check\endcsname{\textcolor{#1}{\ding{51}}}%
}
 \colourcheck{green}
 \newcommand*\colourx[1]{%
  \expandafter\newcommand\csname #1x\endcsname{\textcolor{#1}{\ding{55}}}%
}
 \colourx{red}
 \colourcheck{black}
 \colourx{black}

\makeatletter
\renewcommand{\maketitle}{\bgroup\setlength{\parindent}{0pt}
\begin{flushleft}
  \textbf{\@title}

  \@author
\end{flushleft}\egroup
}
\makeatother

\title{\Huge\bfseries Dynamic virtual ecosystems as a tool for detecting large-scale responses of biodiversity to environmental and land-use change}
\author[1,5]{Claire Harris}
\author[2,5]{Neil Brummitt}
\author[3,5]{Christina A. Cobbold}
\author[4,5]{Richard Reeve\footnote[1]{
Corresponding author: Richard.Reeve@glasgow.ac.uk}}
\affil[1]{Biomathematics and Statistics Scotland, UK}
\affil[2]{Department of Life Sciences, Natural History Museum, London, UK}
\affil[3]{School of Mathematics and Statistics, University of Glasgow, Glasgow, UK}
\affil[4]{School of Biodiversity, One Health and Veterinary Medicine, College of Medical, Veterinary and Life Sciences, University of Glasgow, Glasgow, UK}
\affil[5]{Boyd Orr Centre for Population and Ecosystem Health, University of Glasgow, Glasgow, UK}
\date{}

\titlespacing\section{0pt}{12pt plus 4pt minus 2pt}{0pt plus 2pt minus 2pt}
\titlespacing\subsection{0pt}{12pt plus 4pt minus 2pt}{0pt plus 2pt minus 2pt}
\titlespacing\subsubsection{0pt}{12pt plus 4pt minus 2pt}{0pt plus 2pt minus 2pt}

\newcommand{\beginsupplement}{%
        \setcounter{table}{0}
        \renewcommand{\thetable}{S\arabic{table}}%
        \setcounter{figure}{0}
        \renewcommand{\thefigure}{S\arabic{figure}}%
     }
\begin{document}

%: ------- Chapter name -------
%\begin{titlepage}
       \vspace*{1cm}
 
       \maketitle
 
       \vfill

%\end{titlepage}

%\clearpage

\graphicspath{{figures/PNG/}{figures/PDF/}{figures/}}

%\linenumbers

%: ------- Start sub-document  -------

\section*{Abstract}

%\begin{enumerate}
%    \item 
Ecosystems are governed by dynamic processes such as competition for resources, reproduction and dispersal. These shape their biodiversity and how the system responds to change. Current approaches to modelling ecosystems, especially plants, focus on either describing fine-scale processes for individual species or broad-scale patterns for limited groups of plant functional types.
%    \item 

Digitisation of herbarium and other plant records has unlocked a wealth of information that can be used to drive models of plant communities and make predictions for their future under different scenarios of climate change. The advent of increased computational capacity and fast, high level programming languages allows for simulation of such landscapes at unprecedented scales.
%    \item 

Here, we demonstrate a tool for Ecosystem Simulation through Integrated Species Trait-Environment Modelling (EcoSISTEM), which models plant species across multiple ecosystem sizes, from patches and small islands to regions and entire continents. These simulated ecosystems support the ability to generate many different types of habitat, as well as reproducing different disturbance scenarios such as climate change, habitat loss and invasion. EcoSISTEM also reproduces examples of real-world species distributions by integrating plant occurrence records and global climate reconstructions to simulate plant species throughout the continent of Africa for the past century.
%    \item 

EcoSISTEM allows us to flexibly explore the dynamics of tens of thousands of species interacting across a continent. The code parallelises efficiently across multiple nodes on high performance computing platforms, and has been scaled up to run on over 1000 cores. It allows us to study the impact of changes to climate, resources and habitat and investigate real-life mechanisms surrounding climate change and biodiversity loss.
%\end{enumerate}

\section*{Introduction}

Following the recent Intergovernmental Panel on Climate Change (IPCC) report detailing how human-induced climate change has unequivocally altered our planet, we are now more aware than ever that deep changes need to be made to prevent future climate catastrophes \citep{IPCC2021}. Biodiversity loss, in particular, is expected to increase globally in response to loss of suitable habitats, in turn due to extreme weather events, sea level rise and ever-increasing temperatures. Plant ranges are limited by their dispersal capacities, often more so than animals, and for some species that exist in a particular climate niche, there may be no suitable alternative reachable habitat \citep{Cunze2013}. The International Union for the Conservation of Nature (IUCN) Sampled Red List Index (SRLI) for Plants already estimates that at least one quarter of plants are threatened with extinction, while there are many thousands more species for which we do not have sufficient data to make an assessment \citep{Brummitt2015a}. Great efforts have been undertaken over the past few decades to digitise herbaria and natural history records into online databases such as the Global Biodiversity Information Facility (GBIF) \citep{GBIF2019, Soltis2016}. Such data can be coupled with modelling efforts in order to fill these gaps.

Africa is estimated to host approximately 45,000 plant species in total \citep{Klopper2006} and has been the target of recent efforts to unite multiple datasets of vascular plant records \citep{Dauby2016}, although there are many areas estimated to have a higher species richness than the collection efforts reflect \citep{Sosef2017}. The continent is characterised by extremes of biodiversity, from the vast, species-poor Sahara desert, to the rain forests of the Congo Basin, the second largest rain forest region in the world storing more carbon per land area than the Amazon \citep{Sullivan2017}, to the Cape floristic region, the most species-rich temperate area \citep{Linder2014, Schnitzler2011}. However, due to deforestation and destructive land-use practices, the tropical forests throughout Africa are now estimated to be net sources of carbon \citep{Baccini2017}. It is now more important than ever that we correctly characterise the biodiversity of these large-scale regions, so we can use this as a baseline for future measurements of loss. Advances in computing and big data approaches and the advent of fast, high-level programming languages such as Julia \citep{Bezanson2015} allow for more sophisticated modelling to be carried out much more easily than could ever have been achieved before \citep{Poisot2018}. 

Although the need for large scale, species-focused modelling frameworks is urgent, those with greater geographical and taxonomic scope lack important biological processes such as demographics, dispersal and competition \citep{Urban2016}. Calls have already been made for more general models of biodiversity, highlighting the need for a framework similar to the IPCC to compare large-scale climate models, alongside monitoring of climate variables \citep{Purves2013}. Comparison with the IPCC is apposite, as climate change and land-use change are together the biggest threats to worldwide biodiversity. Dynamic models like the `Madingley' model \citep{Harfoot2014} already show success in replicating biodiversity patterns at broad functional group scale, despite reducing all terrestrial plants to a single group.

\begin{figure}[H]
    \begin{center}
        \includegraphics[height=\vsize/2]{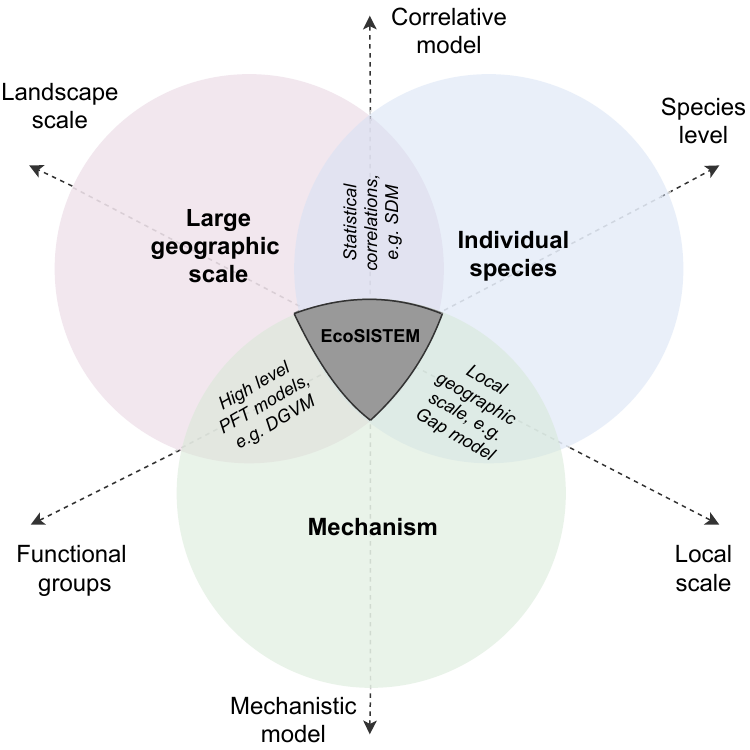}
    \end{center}
    \caption{Schematic characteristics of three predominant approaches for modelling responses of biodiversity to climate change. Each characteristic can be found on an axis: geographic scale spanning from small to large, model approach spanning from mechanistic to correlative, and grouping of individuals from individual species to functional groups. Previous approaches, like dynamic global vegetation models (DGVMs), species distribution models (SDMs) and forest gap models tend to exist in the overlap of at least two of these properties, whereas EcoSISTEM unites all three, to operate over large scales, for individual species undergoing dynamic processes.}
    \label{fig:climatemodels}
\end{figure}

Traditionally there have been three main approaches to modelling biological responses to climate change (Figure \ref{fig:climatemodels}). Species-level approaches determine statistical correlations between current climate and species distributions to project future scenarios of environmental change \citep{Keith2008, Thuiller2008}. Due to their static nature, species distribution or environmental niche models (SDM/ENM) fail to account for important ecological processes affecting distributional change, such as competition, dispersal and evolution \citep{Pearson2003, Akcakaya2006, Thuiller2008, Hannah2014}. Although SDMs are often developed species by species, this approach can be extended to global scope \citep[e.g.][]{Brummitt2021}, but this would require modelling thousands of individual species to reveal community-level compositional change \citep{Hannah2014}. Furthermore, SDMs are often based upon incomplete species records and their interactions with climate or habitat and, without explicitly including biological mechanisms, predict new environments poorly.

Many models do include whole-community dynamics at varying geographic and taxonomic scopes. Dynamic global vegetation models (DGVMs) simulate broad species classes as plant functional types (PFTs), with explicit mechanisms for processes including photosynthesis, growth and carbon cycling. DGVMs can be coupled with Earth system models incorporating feedbacks between environment and vegetation for improved climate predictions. For example, the Joint UK Land and Environment Simulator (JULES) formerly classified plants into only five functional groups: broad-leaved and needle-leaved trees, C3 and C4 grasses, and shrubs \citep{Mercado2011}. Coupled DGVMs are adept at modelling direct and indirect effects of climate change and disturbance on different vegetation types. However, although improvements include more detailed functional groups in JULES \citep{Harper2016}, these coarse PFTs limit the use of DGVMs in conservation planning, or differential within-group responses to changing environments \citep{Thuiller2008}. Equally, such models are primarily forced by detailed historical reconstructions and future climate predictions and often lack direct inputs of anthropogenic disturbance, such as habitat loss or land-use change \citep{Quillet2010}.

At a much finer taxonomic and spatial scale there is the gap model, which simulates colonisation of forest gaps from falling trees. Although dynamic, including processes such as growth and competition, gap models act at the individual tree level to predict spatially-explicit changes to forest composition \citep{Thuiller2008, Hannah2014}. Gap models are often of commercially important tree species in small areas, for which demographic and dispersal data are readily available. They also historically focus on trees of temperate regions, although have been extended in recent years to grasslands, shrubs, and tropical trees \citep{Moorcroft2001, Fischer2016}. Scaling up to regional or continental levels is extremely computationally challenging, notwithstanding progress in simulating tree dynamics for the Amazon rain forest using parallel processing and high performance computing \citep{Shugart2018, Fischer2016}. Indeed gap models of larger areas often rely on aggregating species back into functional groups, similar to their DGVM counterparts \citep{Shugart2018}.

No single approach currently provides all functionality necessary to model biodiversity change at large scales -- a composite approach is needed. Such a model needs to sit in the overlap between DGVMs and traditional single-species models, operate at large scales, for many individual species, and include such biological processes as niche preferences, competition and dispersal (Figure \ref{fig:climatemodels}). Here we present exactly such a composite dynamic ecosystem model -- EcoSISTEM -- and demonstrate its properties through idealised simulations of plant ecosystems across a range of scales and habitat types, from local patches with constant climates, to regional- or continental-scale areas with varied climates. We describe several example scenarios and provide information on the computational efficiency of the code. Finally, we demonstrate the capacity of the model to reconstruct real-world plant ecosystems across entire continents, drawing together plant records from GBIF and global climate reconstructions for the past century. 

\section*{Methods}

\subsection*{General model structure}

The simulator is built around important, related facets of plant community composition, including trade-offs between resource availability and species' requirements, between the environmental variables and niche preferences. EcoSISTEM is a flexible and extensible framework  used to simulate plant dynamics.

\begin{figure}[H]
    \begin{center}
        \includegraphics[width=\hsize]{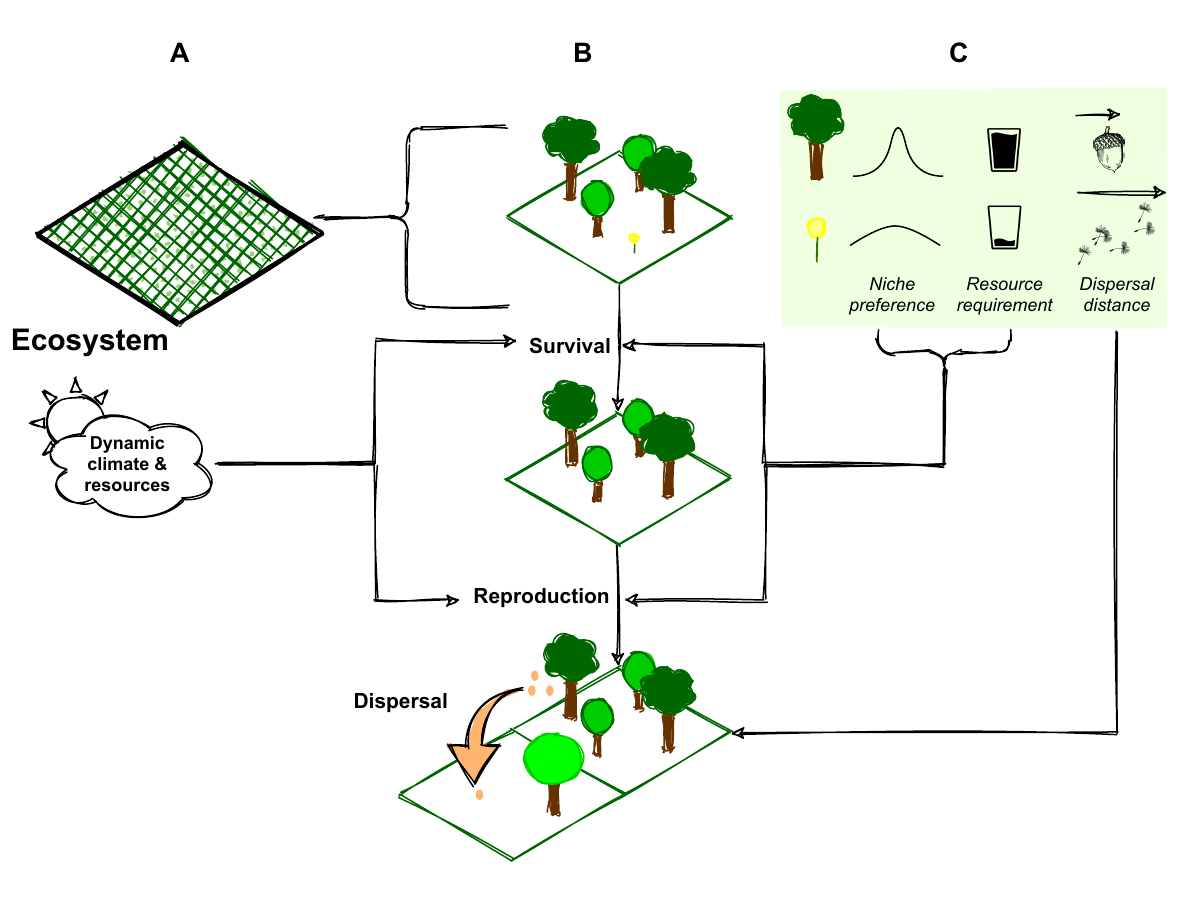}
    \end{center}
    \caption{Overview of the EcoSISTEM model. A) An Ecosystem is made up of a gridded landscape containing plant species with associated climate (e.g. temperature, rainfall) and resource (e.g. sunlight, soil water) layers, which can be updated at each time step. B) In each grid square, species undergo several processes in a single time step. Niche preference and resource competition rules play out across species to determine how many individuals survive to reproduce. Following reproduction, seeds are dispersed across the landscape according to a species-specific dispersal kernel. C) Each species has a set of traits that determines their capacity to compete, reproduce and disperse. Here, the first species has a narrower niche tolerance, greater resource requirement and shorter dispersal distance than the second.}
    \label{fig:modeloverview}
\end{figure}

The 2-D landscape is discretised across a regular grid, and accommodates an arbitrary number of plant species within an ecosystem of any size (we have tested it across Africa with over 50,000 species -- the number of plant species on that continent). Within each grid cell we track the simulated change in abundance of each species over time.
Resource (e.g. sunlight (kW) and water (mm)) availability and environmental data (e.g. temperature ($^{\circ}$C)) are inputs into the model that affect the growth and survival of the plant populations. At each time step (e.g. monthly) the resource and environment is updated, and species' niche preferences, e.g. temperature, and resource requirements are traded off against availability in that grid cell. The match of species to environment and resource needs determines reproduction and death rates. At the end of each time step, after reproduction, seed dispersal takes place. An overview of the model is presented in Figure \ref{fig:modeloverview}.

\subsection*{Biological assumptions}
\label{sec:biolassumptions}
First, we assume that plant dispersal only takes place via seed dispersal, i.e. dispersal follows immediately after seed production, as is observed in most plant species. Here we use a Gaussian dispersal kernel, which concentrates offspring close to the parent plant; however, the model can readily incorporate alternative choices of dispersal kernel. Second, reproduction is modelled as a Poisson process of germination rate per individual (or reproducing unit) and current population size, assuming that time until next seed production is exponentially distributed. This approximates the seed production processes, with an average number of germinations per adult plant per month determined by whether a plant is exposed to environmental conditions suitable for seed production, and whether there are sufficient resources to do so. Therefore, even though reproduction could occur at every time step in the simulation, environmental conditions and resource availability  limit reproduction to windows of suitable conditions and show seasonal reproductive patterns in seasonally-fluctuating climates. Similarly, the probability of death also depends on environmental niche suitability and resource availability, and we model death as a Bernouilli process~\citep{Caswell2001}. These distributions are standard in demographic modelling \citep{Chisholm2014, Melbourne2012}.

\subsection*{Population demographics}

The population dynamics of each plant species in the system are described by equation \ref{eq:overall}. The spatial domain is divided into  $N$  grid cells. The population at the next time step, ${t+\delta t}$, at grid location $X$ is a result of offspring dispersing and adults surviving from the previous time step $t$. The time step $\delta t$ can be altered, but is set at 1 month in simulations. %
%
%\begin{large}
\begin{equation}
\label{eq:overall}
pop_s({t+\delta t}, X) = \underbrace{\vphantom{\sum_{Y=1}^{N} } (pop_s(t, X) - Deaths_s(t, X))}_{\text{\footnotesize \# of adults surviving}} + \underbrace{\sum_{Y=1}^{N} Seeds_s(t, Y) \times k_s(X, Y)}_{\text{\footnotesize Dispersal of seeds from Y to X}}
\end{equation}
%\end{large}%
%
The number of individuals of species $s$ at time $t$ in location $X$ is  $pop_s(t, X)$, $Seeds_s(t, X)$ is the number of seeds produced by species $s$ and $Deaths_s(t, X)$ is the corresponding number of deaths. Finally, $k_s(X, Y)$ is the probability that a seed produced by a plant of species $s$ at location $Y$ disperses to $X$ to grow.

\subsubsection*{Reproduction and death}

For each plant species, the number of individuals dying and the number of seeds produced at each time step is drawn from a Poisson and a binomial distribution, respectively, with the expected number of seeds per adult plant and probability of death during a time period of $\delta t$ as $R_{b,s}$ and $1-\exp(-P_{d,s})$, respectively (Equations \ref{eq:births} and \ref{eq:deaths}).%
%\begin{large}
\begin{equation}
\label{eq:births}
Seeds_s(t, X) \sim Poisson(pop_s(t, X) \times R_{b,s}(t, X)),
\end{equation}
\begin{equation}
\label{eq:deaths}
Deaths_s(t, X) \sim Binomial(pop_s(t, X), 1-\exp(-P_{d,s}(t, X))),
\end{equation}
%\end{large}%
%
where $R_{b,s}$ and $P_{d,s}$ depend on expected rates of seed production and death rates, $b_s$ and $d_s$, in a suitable environment with adequate resources, moderated by the resource availability and environmental suitability via $B_{e,s}$ and $D_{e,s}$:

\begin{equation}
\label{eq:ratebirth}
 R_{b,s}(t, X) = b_s \times B_{e,s}(t, X) \times \delta t, 
\end{equation}
\begin{equation}
\label{eq:probdeath}
 P_{d,s}(t, X) = d_s \times D_{e,s}(t, X) \times \delta t.
\end{equation}

In the expressions for $B_{e,s}$ and $D_{e,s}$ (Equations \ref{eq:birthenergy} and \ref{eq:deathenergy}, respectively), the last terms describe the effects of resource competition on each species, which depends on the ratios of resource $i$, $K_i(t,X)$, to the total of resource $i$ used by all individuals of all species  in the grid cell, $E_i(t, X)$  ($= \sum_{s = 1}^{S}{\epsilon_{s,i} \times pop_s(t, X)}$, where $s$ denotes the species and $\epsilon_{s,i}$ is the $i$-th resource requirement of species $s$). 
Here we assume there is a total of $S$ species and $R$ resources.
Resource $K_i(X,t)$ is determined from resource (e.g. sunlight, water, etc.) data input into the model and is not affected by population abundance. There are no advantages when more resource is available than individuals to consume it. 

\begin{figure}[H]
    \begin{center}
        \includegraphics[width=\hsize]{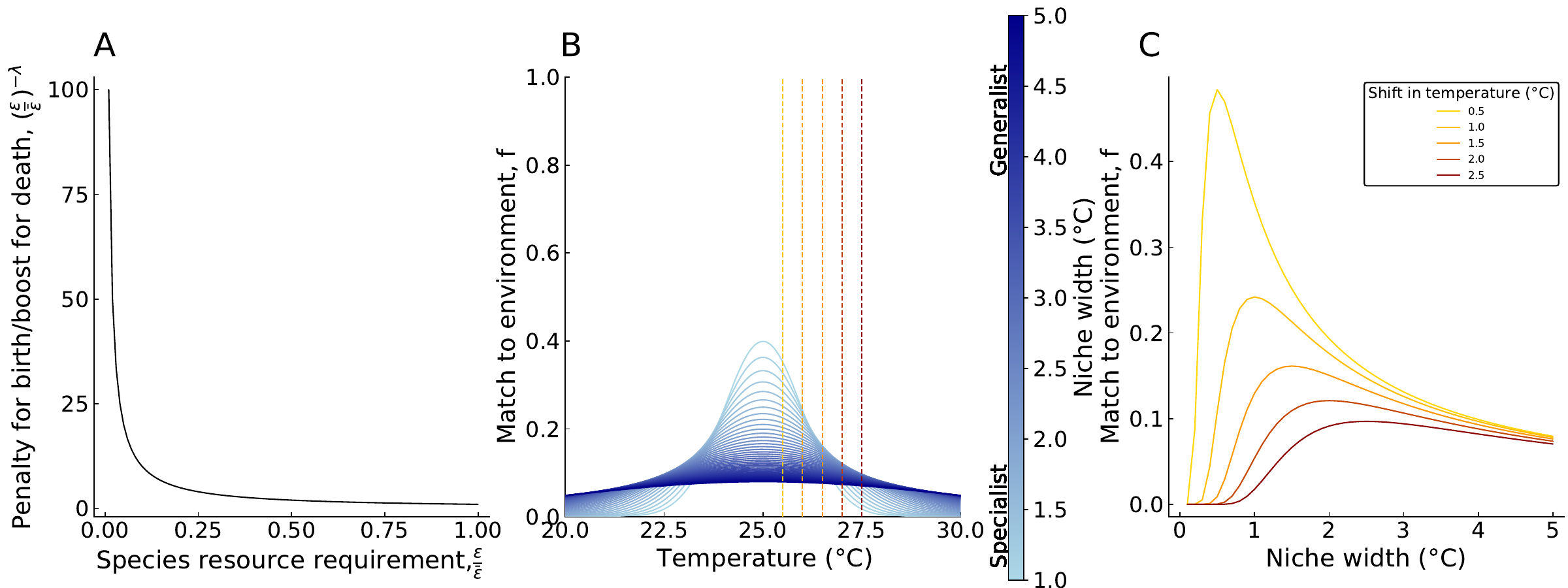}
    \end{center}
    \caption{A) Resource competition: The relationship of species' resource requirements to reproduction and death processes. Species with larger requirements have a slower life cycle, produce seed less often and live longer than those with smaller requirements. $\lambda$ is assumed to be 1; B) Niche preference: Match to the environment ($f$) for example species with a temperature preference of 25$^\circ$C, across an environmental temperature range of 20 - 30$^\circ$C for different niche widths (blue lines). The dashed and dotted black lines show the match to the environment for species at their niche optimum and 1$^\circ$C outside it, respectively; C) Environmental change: Match to the environment ($f$) for example species with different niche widths for an environment 0.5 - 2.5$^\circ$C outside the species niche optimum.}
    \label{fig:nichewidth}
\end{figure}

Species requiring high levels of resources reproduce less often, but also live longer, as observed in many  species including plants \citep{Reich2001} (Figure \ref{fig:nichewidth}A), and reflected in the first terms in Equations \ref{eq:birthenergy} and \ref{eq:deathenergy}. This process is also modulated by fitness of a species in its environment (second terms in Equations \ref{eq:birthenergy} and \ref{eq:deathenergy}), with those more mismatched facing decreased ability to reproduce and higher mortality through adjusted resource requirements, $\chi_{i,s}$.
For $R$ resources and $1$ environmental variable, then combined with resource competition, the above dynamics are captured in Equations \ref{eq:birthenergy} and \ref{eq:deathenergy}.

\begin{equation}
\label{eq:birthenergy}
 B_{e,s}(t, X) = \underbrace{\vphantom{\min\left(\frac {K(X)} {E(t, X)}, 1\right)} \left(\frac{\epsilon_{1,s}}{\bar\epsilon_1}\times\dots\times\frac{\epsilon_{R,s}}{\bar\epsilon_R}\right)^{-\lambda}}_{\substack{Relative \: resource \\ requirement}} \times \underbrace{\vphantom{\min\left(\frac {K_1(t,X)} {E_1(t, X)},1\right)}(\chi_{1,s}\times\dots\times\chi_{R,s})^{-\tau}}_{\substack{Environment \\ suitability }} \times \underbrace{\min\left(\frac {K_1(X)} {E_1(t, X)},\dots, \frac {K_R(t,X)} {E_R(t,X)}, 1\right)}_{\substack{Resource \\ competition}},
\end{equation}

\begin{equation}
\label{eq:deathenergy}
 D_{e,s}(t, X) = \underbrace{\vphantom{\frac {E(t, X)} {K(X)}} \left(\frac{\epsilon_{1,s}}{\bar\epsilon_1}\times\dots\times\frac{\epsilon_{R,s}}{\bar\epsilon_R}\right)^{-\lambda}}_{\substack{Relative \: resource \\ requirement}} \times \underbrace{\vphantom{\frac {E(t, X)} {K(X)}} (\chi_{1,s}\times\dots\times\chi_{R,s}))^{\tau}}_{\substack{Environment  \\ suitability}} \times \underbrace{\max\left(\frac {K_1(X)} {E_1(t, X)},\dots, \frac {K_R(t,X)} {E_R(t,X)}\right)}_{\substack{Resource \\ competition}},
\end{equation}
where the average resource requirement of resource $i$ for all species is $\bar\epsilon_i$ ($ = \frac{1}{S}\sum_{s=1}^{S}{\epsilon_{s,i}}$), and $\frac{\epsilon_{s,i}}{\bar\epsilon_i}$  denotes the relative resource $i$ requirements of species $s$. The extent to which the species environmental niche ($\bar{T}_s$) matches the experienced environment ($T$) is  captured in $\chi_{i,x}$ in Equation \ref{resourcereq}. The relative resource $i$ requirement of species $s$, $\frac{\epsilon_{i,s}}{\bar\epsilon_i}$ is adjusted by the fit of the species niche preference to the current environment, $f(T;\bar{T}_s)$ (Equation \ref{resourcefunc}), to penalise reproduction and increase mortality for species outside their natural habitats. $f$ is modelled here as a Gaussian curve so that the penalty for being far away from the niche preference ($\bar{T}_s$) is normally distributed \citep{DeBlasio2015}, examples for which are given in Figure \ref{fig:nichewidth}B. However, this can easily be modified to consider alternative distributions.

\begin{equation}
\label{resourcereq}
\chi_{i,s} = \frac {\epsilon_{i,s}} {\bar\epsilon_i \times f(T;\bar{T}_s)},
\end{equation}
where
\begin{equation}
\label{resourcefunc}
f(T;\bar{T}_s) =  \frac {1} {\sqrt{2\pi\psi_s}} \times e^{\frac{-(\bar{T}_s - T)^2}{2\psi_s^2}},
\end{equation}
and $T$ is the current environment in grid cell $X$, $\bar{T}_s$ is the niche mean for species $s$ and $\psi_s$ the niche width (standard deviation) of the species. This rewards species staying within their niche more if the niche is narrow, thus giving an advantage to specialists over generalists.

%$\lambda$ and $\tau$ are fixed parameters for the %importance of resource requirements and trait-environment relationships on reproduction-death processes.  In the absence of data to fit these parameters, for  simulations presented here they were tuned to give plausible results: $\lambda =1$ and $\tau = 0.001$.

\subsubsection*{Dispersal}

\begin{figure}[H]
    \begin{center}
        \includegraphics[width=\hsize]{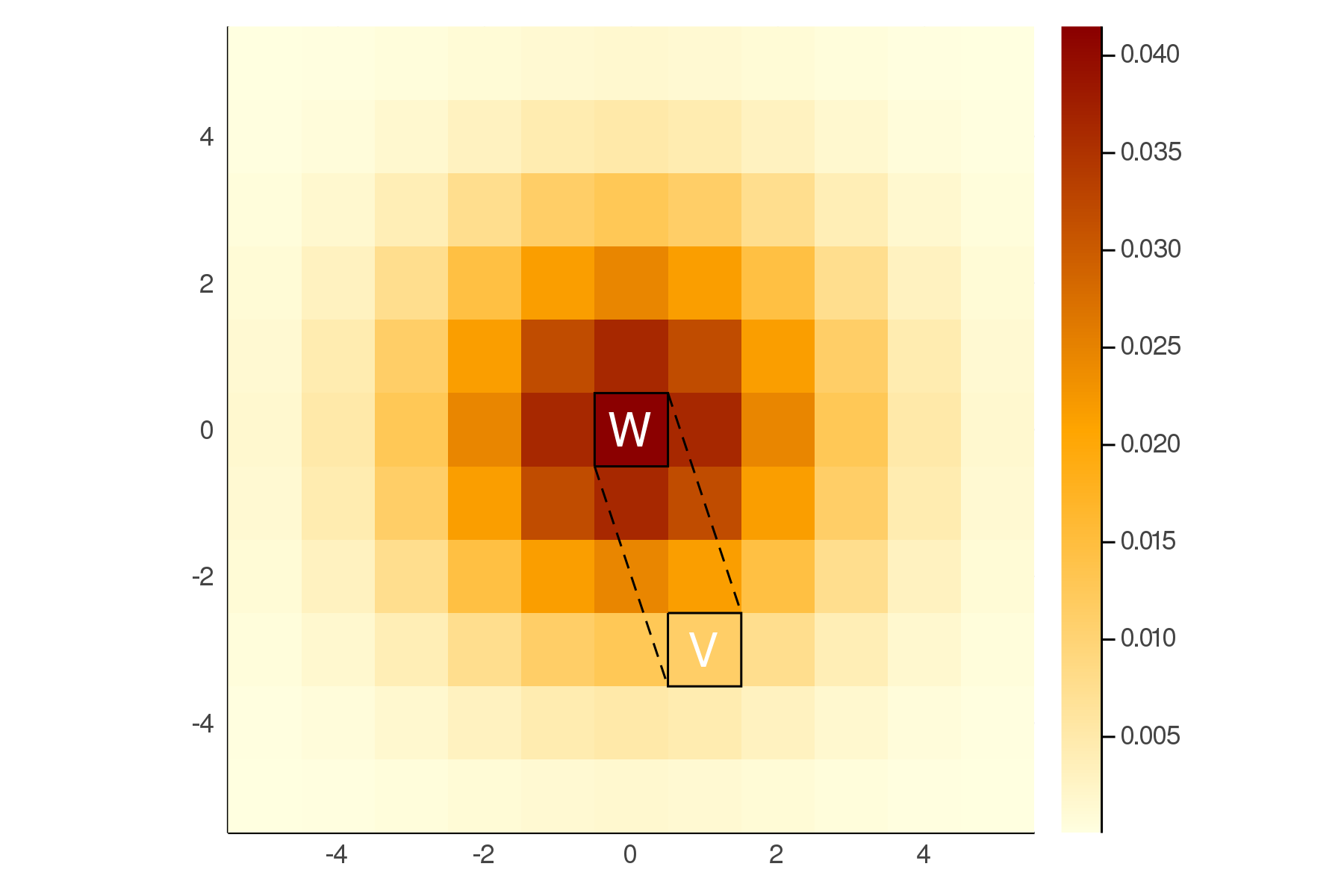}
    \end{center}
    \caption{Diagram of area-to-area dispersal used in EcoSISTEM between source and destination grid squares, $W$ ($x_W, y_W$) and $V$ ($x_V, y_V$),  respectively. This diagram has been adapted from \cite{Chipperfield2011}, in which the probability of moving from $W$ to $V$ is a result of integrating over all possible points in the source cell to all possible destination points in the destination cell. Intensity of colour represents the probability of dispersal.}
    \label{fig:dispersal}
\end{figure}

Only seeds produced in the current time step can disperse. We use a Gaussian dispersal kernel with standard deviation $\sigma_s$. The probability of a seed of species $s$ dispersing from anywhere in the source grid cell, $V$, to destination grid cell, $W$, is $k_s(V, W)$ as described by Equation \ref{eq:gaussdispersal}:

\begin{equation}
\begin{aligned}
\label{eq:gaussdispersal}
k_s(V, \; W ) =  \iiiint \limits_{\substack{
x_V - \frac{1}{2}\ \le\ x\ <\ x_V +  \frac{1}{2}\\
y_V -  \frac{1}{2}\ \le\ y\ <\ y_V +  \frac{1}{2}\\
x_W -  \frac{1}{2}\ \le\ x^\prime\ <\ x_W +  \frac{1}{2}\\
y_W -  \frac{1}{2}\ \le\ y^\prime\ <\ y_W +  \frac{1}{2}}} \dfrac{1}{\pi {\theta^2}}e^{-\dfrac{{(x-x^\prime)}^2+{(y-y^\prime)}^2}{\theta_s^2}} \; dx \; dy \; dx^\prime  \; dy^\prime
\end{aligned},
\end{equation}
where $(x_V, y_V), (x_W, y_W)$ are the grid square indices of the top right corner of the destination and source grid square, respectively.  Space is scaled by grid cell width,  $\eta$, so that in the scaled space grid cells have unit width and height, as illustrated in Figure \ref{fig:dispersal}, and the mean dispersal distance of the species is  $\theta_s= \frac{\sigma_s}{\eta}$.

The use of Gaussian dispersal kernels is contentious \citep{Bullock2017}, but it remains the simplest option in the absence of direct information about dispersal, such as dispersal modes and seed size. We illustrate results with the Gaussian kernel, but alternative dispersal kernels can be easily implemented in EcoSISTEM.

\subsection*{Code base}

\begin{figure}
    \centering
    \includegraphics[width=\hsize]{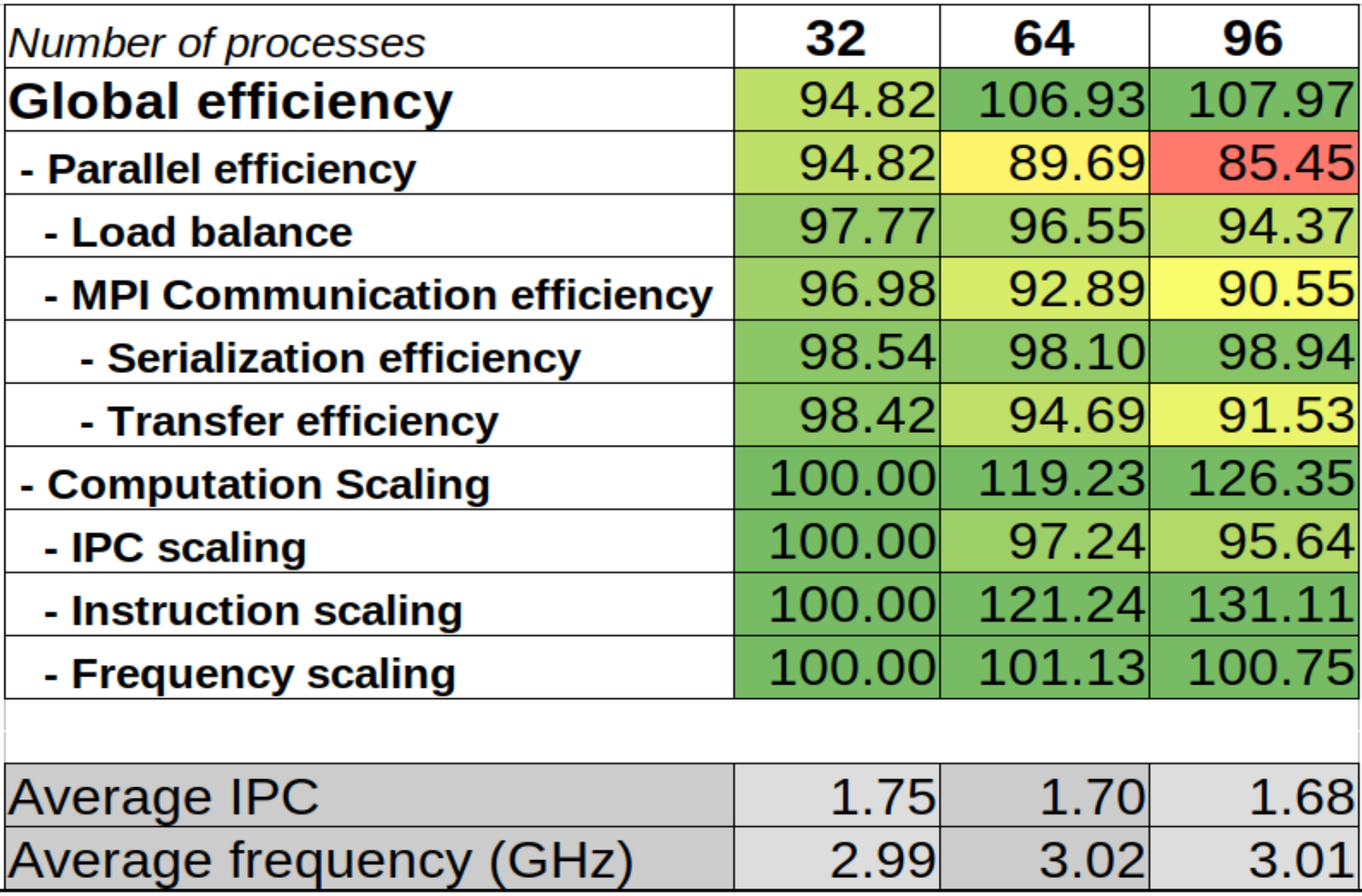}
    \caption{Testing scalability of EcoSISTEM code using POP metrics: EU Horizon 2020 POP Centre of Excellence, POP Standard Metrics for Performance Analysis of Hybrid Parallel Applications -- https://pop-coe.eu/further-information/learning-material/pop-standard-hybrid-metrics-for-parallel-performance-analysis -- shows EcoSISTEM is between 95\% and 108\% efficient as it paralellises across multiple processes.}
    \label{fig:nag}
\end{figure}

All the code and underlying functionality for EcoSISTEM was written in Julia v1.5.3 \citep{Bezanson2015, Besancon2021}. The package is designed to interact with EcoJulia architecture and packages such as \textit{Diversity} and \textit{Phylo} \citep{Reeve2021a, Reeve2021b}, enabling the user to calculate diversity metrics for ecosystems and incorporate phylogenetic relationships between species. The code has been extensively tested and refined for speed and efficiency. It is multithreaded to run efficiently on multi-core workstations, and can be distributed across multiple nodes on high performance computing (HPC) platforms using the Message-Passing Interface (MPI) for inter-process communication \citep{MPI2015}. Testing by the Numerical Analysis Group (NAG, Oxford, UK) on HPC found an essentially perfect (100\%) global efficiency of the parallelisation in tests of up to 96 processes (Figure \ref{fig:nag}), and the code has been run on over 1000 cores.

\subsection*{Data}

For the simulation of Africa, we used the European Centre for Medium-Range Weather Forecasting (ECMWF) climate reanalysis datasets \citep{Laloyaux2018, Dee2011} to form the climate part of the ecosystem, combining the coarser 120km CERA-20C data (1901-2000) with finer scale ERA-interim data (1979-2018) to run simulations at an 80 km resolution. We used minimum monthly temperature ($^\circ$C) and rainfall (mm) estimations for the full timespan, 1901-2018, to describe the climate niche. We also used the same combined ECMWF reanalysis datasets for the resources in the ecosystem. Here, they are surface solar radiation (kW/m$^2$) and soil water volume (m$^3$ water/m$^2$) at layer 1 (up to 7cm depth). 

We filtered out those species with only a single recording point in the underlying GBIF data as they were too data deficient, leaving a total of 39,158 species recorded at least twice in Africa. The climate preferences were extracted from GBIF and ECMWF reanalysis datasets at the time and location of the GBIF record, corrected by sampling effort at that spatial location \citep{Harris2021d}. We took the mean and standard deviation of the temperature and rainfall preferences as the niche mean and niche width for each species. There is little available information on requirements for solar energy and water for most plant species, so we took the environmental conditions they were exposed to as a proxy, assuming that they used all available resource in the areas they were found. Thus, we used the extracted information on resource needs for soil water volume and surface solar radiation, corrected for effort as in the niche preferences. Finally, we used randomised reproduction, mortality and dispersal parameters for this example as this information is unavailable for most plant species.

\subsection*{Experimental design}

To validate the model, simulations of small-scale ecosystem patches and islands were conducted. These were equal in size and seeded with 100 species / 100km$^2$, but boundary conditions differed: islands had hard boundaries that species could not disperse beyond, whereas patches were toroidal and species could disperse in all directions. We set up each example with constant background temperature and ran each simulation for 10 years. We then expanded the case study to a grid with 80km squares and a total area similar to that of a continent. Using this continental backdrop, we investigated how quickly species could colonise large areas, whether they were previously populated or not.

In order to initially seed the Africa simulations, we began by filling the ecosystems with the overall summed counts of each GBIF record according to the locations at which they had been previously found. These counts were adjusted by the availability of resource in each grid square, so that the ecosystem began with a number of individuals close to the equilibrium. Then we ran the simulations backwards in time from 2018 to 1901, adding in each species record at each location at the date it was recorded. This enabled us to simulate a baseline for plant diversity in 1901 from which to start the simulations, which could not be gained from a burn in forward through time. The number of individuals of each species added at each time point was scaled by the inverse of collection effort. This correction ensured that both under- and over-collected areas were seeded with similar numbers of plants despite dramatically different sampling efforts. Once the simulations reached 1901, they were cycled through this same year 10 times in order to ensure they were fully equilibrated. From this starting point, we ran the ecosystems ran through the full monthly climate history through to 2018. We performed 10 stochastic realisations, which was chosen after considering the computational expense of the simulations. 

\section*{Results}
\subsection*{Island, patch and continental dynamics}
\label{sec:modeltest}

\begin{figure}[H]
    \includegraphics[width=\textwidth]{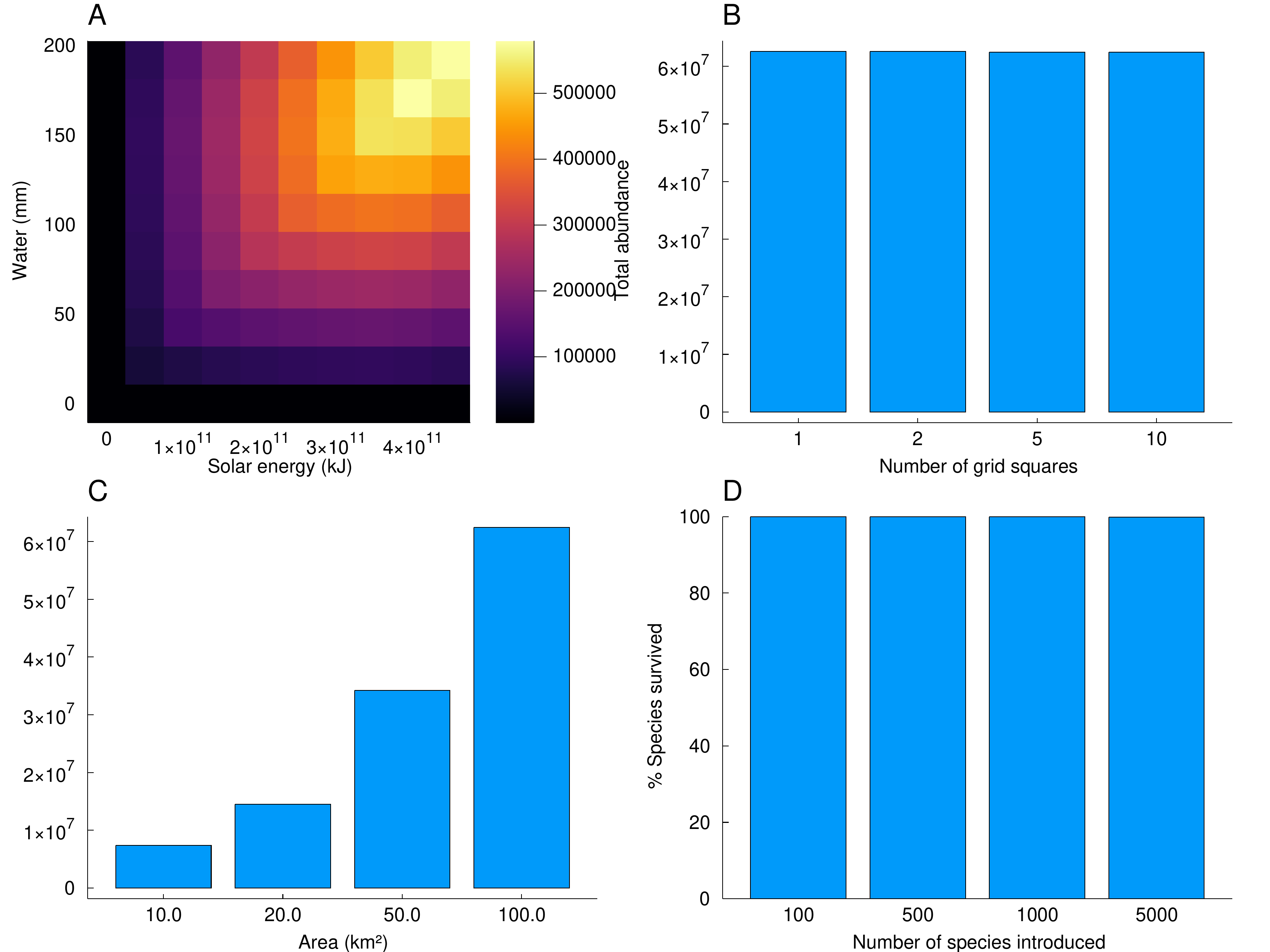}
    \caption{Model testing on island ecosystems. (A) Total abundance of 100 species, with varying resources of water and solar energy across the grid. (B) Total abundance of 100 species, with increasing area size. (C) Total abundance of 100 species, with increasing grid square resolution. (D) Percentage of species that survived after 10 years of simulation.}
    \label{fig:testing}
\end{figure}

\begin{figure}[H]
    \includegraphics[width=\textwidth]{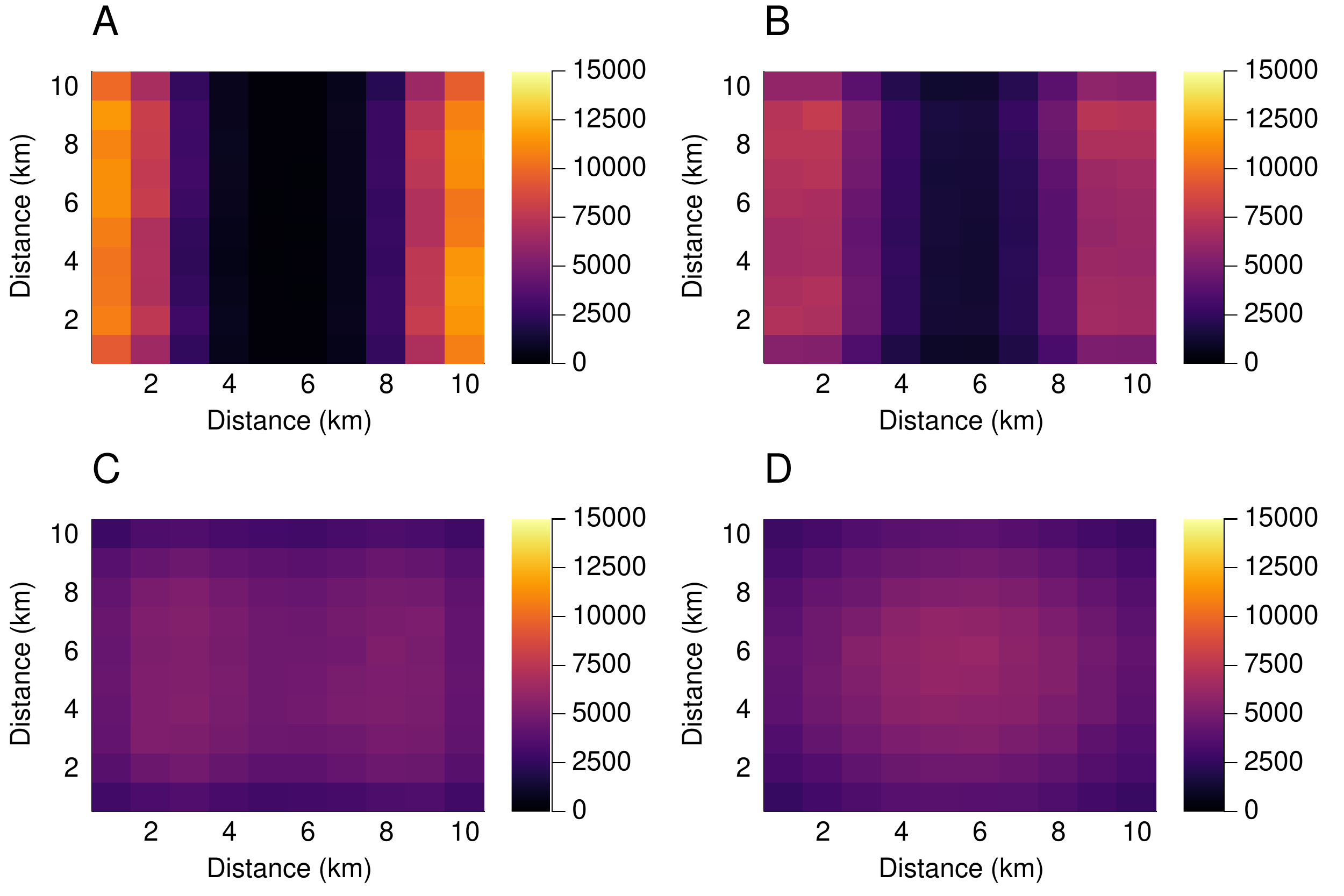}
    \caption{Total abundance of two species in island ecosystems after 10 years of simulation, with species seeded at opposite sides of the island. Those with higher dispersal distances moved further away from their starting populations at a faster rate. (A) Mean dispersal distance of 0.5km, (B) mean dispersal distance of 1km, (C) Mean dispersal distance of 2km, (D) Mean dispersal distance of 4km}
    \label{fig:dispersalSD}
\end{figure}

We confirmed that abundance depends on available resources (Figure \ref{fig:testing}A), here on an island ecosystem with two resources, water and sunlight, each on a gradient West to East and South to North, respectively. All species were seeded with the same resource requirements and vital rates. Abundance increased in cells with more water and sunlight, with some edge effects. As expected, abundances were invariant to the grid resolution (Figure \ref{fig:testing}B), and ecosystems with greater areas could also support more individuals (Figure \ref{fig:testing}C). We also tested ecosystems seeded with varying numbers of species, and similar numbers survive until the simulation ends in each case (Figure \ref{fig:testing}D). Figure \ref{fig:dispersalSD} shows overall abundance on an island with two species at opposite sides of the island after ten years of simulation: species with higher average dispersal distances moved faster and further into the unpopulated centre.

\begin{figure}[H]
    \includegraphics[width=\textwidth]{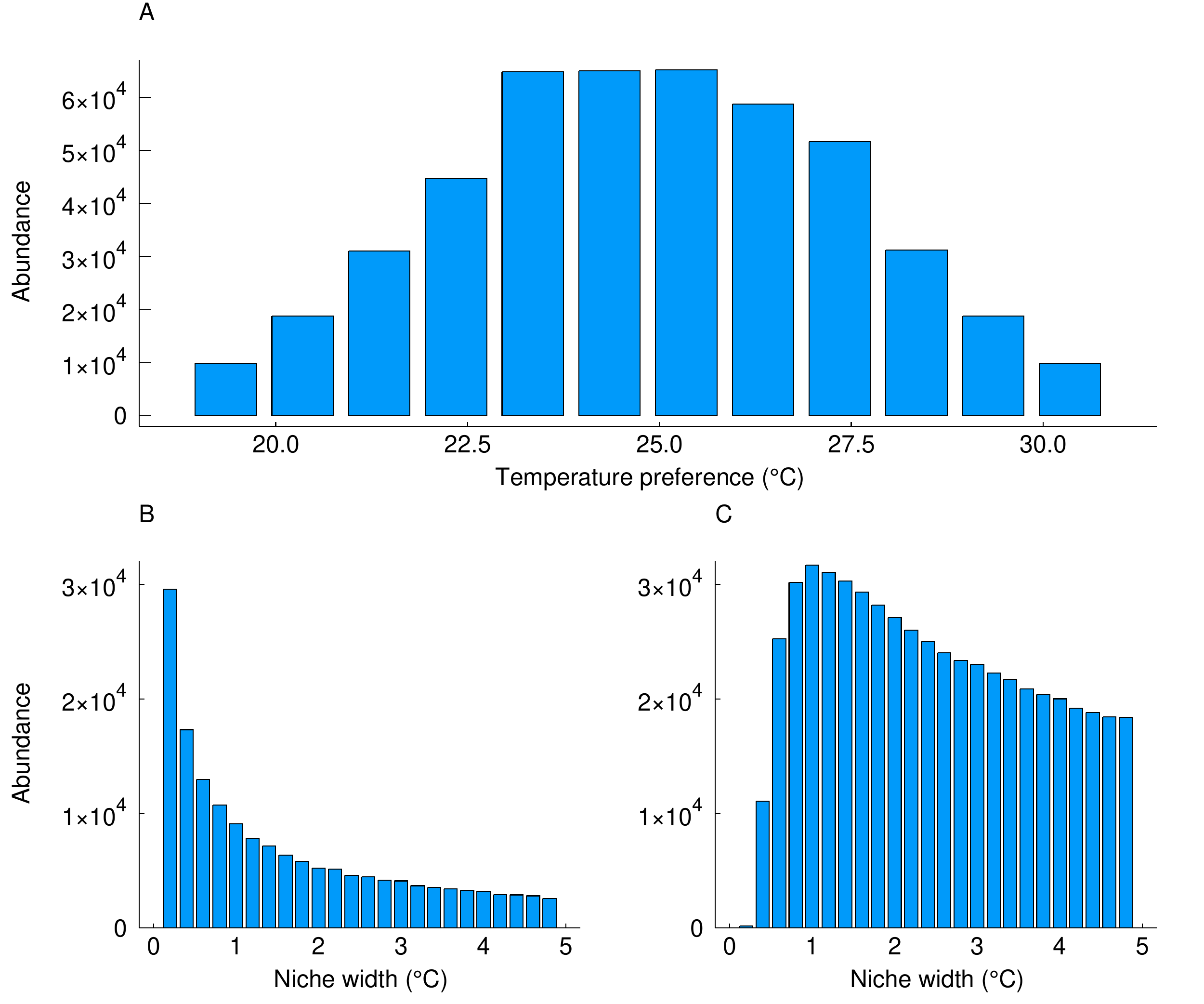}
    \caption{Abundance of species across a 100km$^2$ patch ecosystem with 100 species, (A) with different temperature preferences and a homogeneous climate of 25$^{\circ}$C, (B)  with different niche widths and a temperature preference for 25$^{\circ}$C and (C) different niche widths with a homogeneous climate shifted to 26$^{\circ}$C.}
    \label{fig:diffopts}
\end{figure}

We tested whether individual species' niche preferences and widths functioned as expected, on a small patch with species given different niche preferences and all other parameters kept equal. When all species had equal niche widths, species with niche preferences nearer the 25$^{\circ}$C optimum were more abundant (Figure \ref{fig:diffopts}A); when all species had a preference for 25$^{\circ}$C and varying niche widths, generalists with broader niches were less abundant than specialists with narrow niches (Figure \ref{fig:diffopts}B). This is expected given the decline in species' match to their optimum environment as niche width increases in Figure \ref{fig:nichewidth}B. However, if temperature was then increased by 1$^{\circ}$C, species with narrowest niches went extinct, and generalists became more abundant (Figure \ref{fig:diffopts}C), with a preference for species with a niche width of around 1$^{\circ}$C, as predicted in Figure \ref{fig:nichewidth}D.

\begin{figure}[H]
    \begin{center}
        \includegraphics[width=\textwidth]{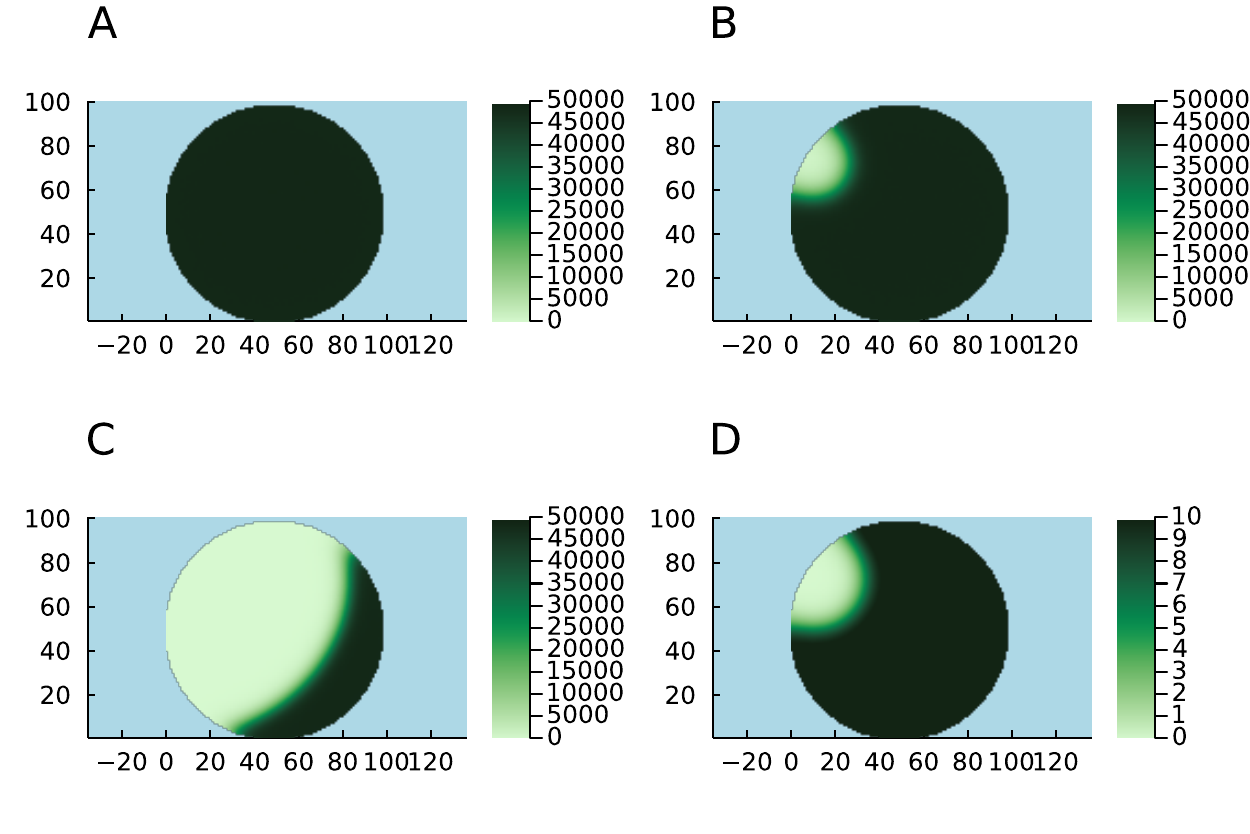}
    \end{center}
    \caption{100 year simulations of an example circular ``continent'' 8,000km across with 50,000 species. (A) Species richness after 100 years of simulation with all species equal. (B) Species richness after 50 years, with one specialist introduced. (C) Species richness after 100 years, with one specialist introduced. (D) Shannon entropy after 50 years with one specialist introduced (large values indicate more evenness).}
    \label{fig:africa}
\end{figure}

EcoSISTEM was designed to scale to much larger areas, supporting many more species. To illustrate this, we simulated 50,000 plant species at an 80km grid scale over a circular ``continent'' 8,000km across, with a constant environment of 25$^{\circ}$C. When all species have equal fitness in the habitat, all 50,000 co-exist over long time scales of over 100 years (Figure \ref{fig:africa}A).

\begin{figure}[H]
    \begin{center}
        \includegraphics[width=0.5\textwidth]{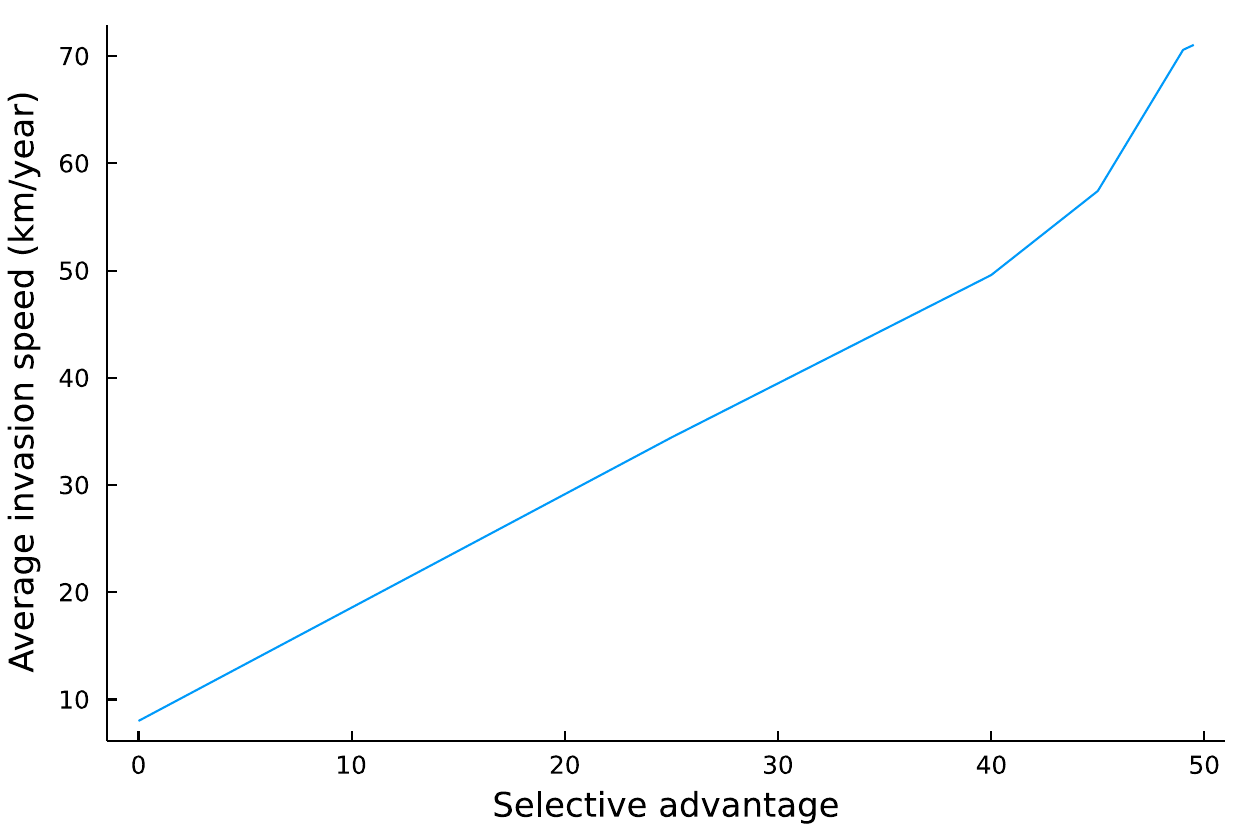}
    \end{center}
    \caption{Invasive capacity of a invasive plant species versus a generalist. Selective advantage is the difference in niche width between the invasive and generalist, and invasion speed is calculated as the average distance travelled per year by the invasive.}
    \label{fig:invasion}
\end{figure}

We also explored the selective advantage of invasive species: introducing an invasive with a narrow niche width tuned to the environment into an continent-sized landscape with an existing generalist, the invasive out-competes the generalist and spreads across the continent. Here, we define selective advantage as the difference in niche width between the invasive and generalist. The larger the invasive's selective advantage, the faster it will invade and colonise the landscape (Figure \ref{fig:invasion}). When we simulate 50,000 generalist species coexisting across a landscape and introduce a single competitive invasive, the invasive can out-compete the local established populations (Figure \ref{fig:africa}B-D). Figure \ref{fig:africa}D shows the Shannon entropy of communities within individual grid squares, using the \textit{Diversity} package \citep{Reeve2021b}. The invasion centre-point, where the invasive dominates, is the most uneven species distribution, with a single species dominating whereas outside, every species coexists. The strength of competitive interaction can be adjusted so that invasives and generalists coexist in the system.

\subsection*{Plant biodiversity of Africa}

\begin{figure}[H]
\begin{center}
\includegraphics[height=0.4\vsize]{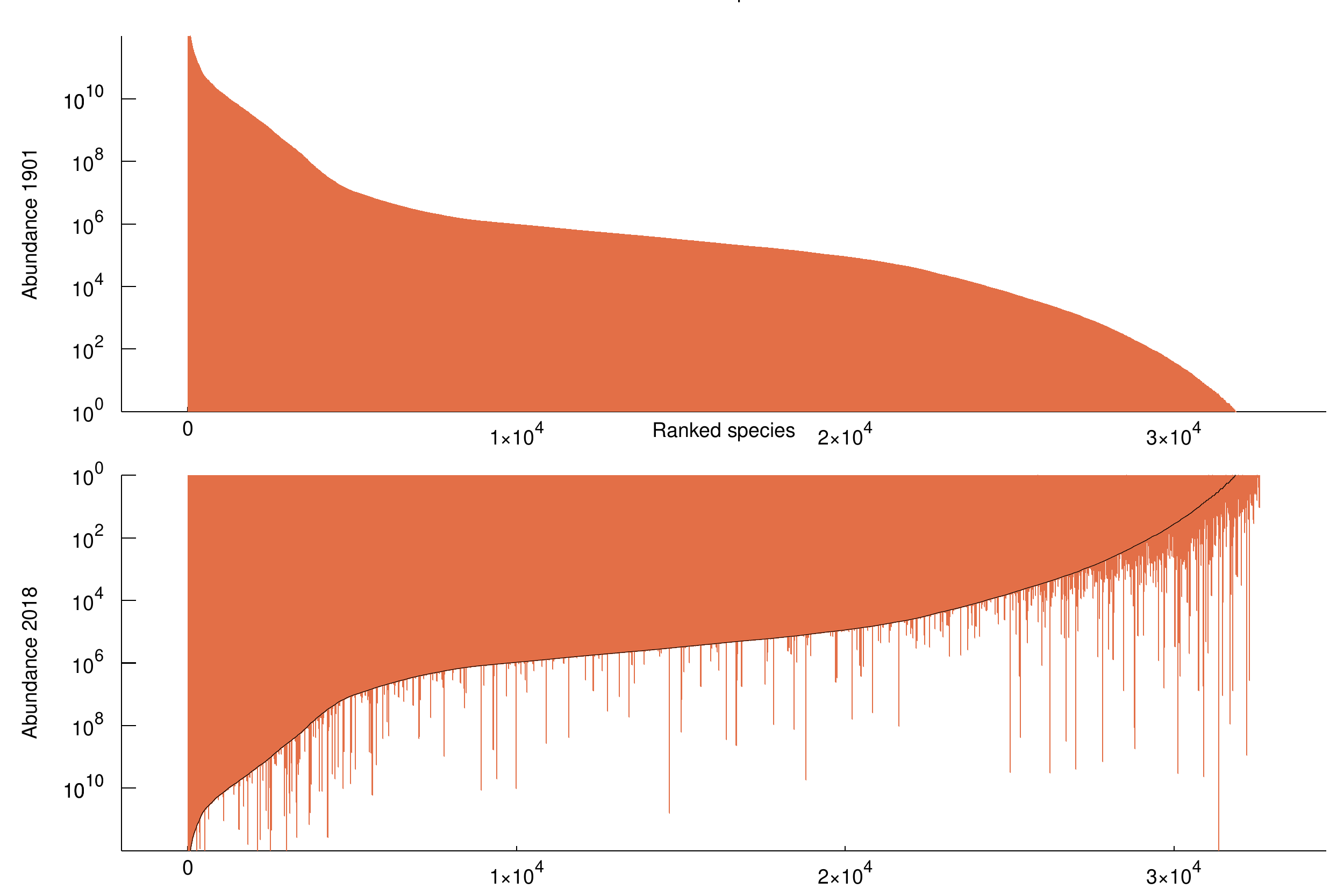}
\end{center}
\caption{Logged ranked abundance of species in the African ecosystem at year 1901 and 2018 for the climate model. Species rank in 2018 is the same as for 1901, and the black line represents the 1901 values for each species.}
\label{fig:climateabun}
\end{figure}

Finally, we simulated the competition, reproduction and dispersal of all of the plant species recorded on the African continent over the 20th and early 21st century, based upon digitised plant records and global climate reconstructions. As a result of the sparsity of GBIF records for plants in Africa many plant species had very low numbers of records, and competition between species during the burnin stage resulted in the survival of only approximately 33,000 species. Preliminary analyses seeding a threshold number of individuals from each species into the environment according to their climate niches suggests that the system can support the full complement of species, but that is not supported by the GBIF data, so we did not pursue it further. Instead we studied the 33,000 surviving species in our simulation study rather than the original 39,158 species. A comparison of their abundance between the beginning and end of the simulation (Figure \ref{fig:climateabun}) suggests that they are relatively stable throughout the course of the simulation, with some stochastic gains and losses for rarer species saw similar losses (on average 1,968 (sd = $\pm1452$) species went extinct).

\begin{figure}[H]
\begin{center}
\includegraphics[width=\hsize]{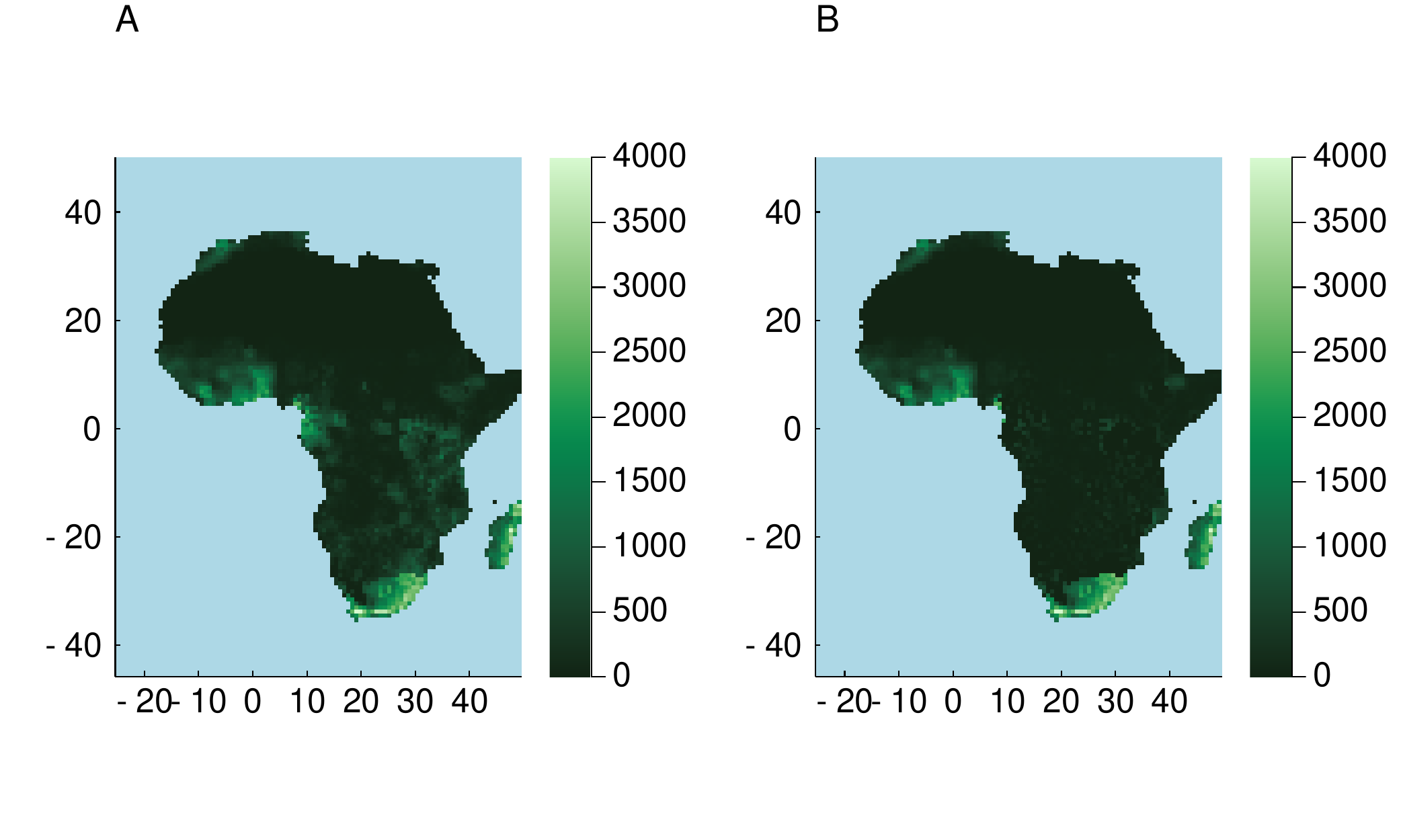}
\end{center}
\caption{Average species richness for the simulation of African plants throughout the course of the timespan, A) 1901. B) 2018.}
\label{fig:SRA}
\end{figure}

Overall, simulated species richness decreased; however, local species richness was concentrated around well-known biodiversity hotspots in the central band of Africa, as well as the South (Figure \ref{fig:SRA}). There were also some species rich regions that surrounded what are in reality urban areas, such as Lagos in Nigeria.

\section*{Discussion}

Digitised taxonomic records from herbaria and natural history collections around the globe offer a unique opportunity, not just to monitor our current and historical plant biodiversity \citep{Harris2021b}, but also to make predictions for the future. In order for this to be feasible we must have models that operate over large geographic scales, at the species-level, that can incorporate a variety of mechanisms essential to plant growth, survival and reproduction. EcoSISTEM is such a tool, capable of simulating dynamics of tens of thousands of plant species at continental or even planetary scales.

The initial case studies, `Island, patch and continental dynamics', show that this model behaves as expected in various scenarios from small ecosystem patches to entire continents. On smaller-scale island and patch habitats, the simulator comfortably represents all desired properties from our (albeit non-exhaustive) list, and the code is tested automatically against these behaviours with an integrated test suite during development. Scaled up to whole continents, the simulator can sustain 50,000 species and demonstrates that an invasive can colonise such a habitat through competitive advantage.

 The main simulation study, `Plant biodiversity of Africa', demonstrates the size and scale of modelling that can be achieved with EcoSISTEM. More complex environments and scenarios of environmental and biotic change can be readily simulated, together with the ability to incorporate climatic data layers, for example from WorldClim or ECMWF. Using such information, we were able to recreate patterns of species richness across the African continent, tracking the course of climate change across the past century. Here, we play out competitive, invasive dynamics across the African continent, comfortably supporting 33,000 species. The strength of the competitive interaction term can be tuned to simulate different propensities for co-existence. All species in the simulation were parameterised from analyses of online natural history records, but were also partly synthetic due to the absence of easily accessible data on reproduction strategies, dispersal and longevity. However, the model can be easily enhanced with any such data from real-life species as it becomes available. Despite the partly synthetic nature of our simulations, however, the model captured many features of estimated species distributions in Africa. Natural history collections offer a unique opportunity to further parameterise these life history traits, especially with the advent of digitised taxonomic descriptions, specimen records and photographs \citep{Harris2021c}. Approaches to extract such information are currently underway, adapting natural language processing techniques to taxonomic descriptions. Future versions of the code will also seek to include important positive interactions between species, such as facilitation \citep{Cavieres2014}.

As there are little data on diversity at this geographic scale, we compared patterns in species richness against those mapped globally \citep{Kier2005,Barthlott2005, Barthlott2007, Brummitt2021} and for Africa \citep{Dauby2016, Linder2014, Harris2021b, Harris2021c}. The species richness exhibited by the models broadly corresponds with well-known biodiversity hotspots. In Africa, the Cape floristic region showed as particularly diverse \citep{Schnitzler2011, Schnitzler2012}, as well as peaks in previously described highly rich regions of rain forest in Central Africa \citep{Linder2014,Barthlott2005}. That the models broadly agree with global species richness patterns supports theories surrounding species richness gradients and climate, since these are the elements currently captured most faithfully by the model. There is thought to be a strong correlation between species diversity and climate when there is one limiting factor involved, such as energy or water availability \citep{Barthlott2005}. The loss of biodiversity, about 5\% of species, observed in simulations over the course of the century is impossible to verify given the absence of cross-sectional data that might inform us about current or historic African biodiversity at anything other than a very small geographic (and often taxonomic) scale.

It is also difficult to set a baseline of climate stability, given that we cannot untangle the Earth's climate history from anthropogenic changes. Here, we chose to repeat the first year of climate reanalysis, 1901, as a proxy of pre-industrial climates. However, there are also evidently more factors involved in the current distribution of plants than climate. For example, urban areas and those undergoing land-use changes, such as those used for agriculture or forestry, are not accounted for in this model, but have a strong impact on the distribution and richness of plant species \citep{Gerstner2014, Phillips2017, Kier2009}. Indeed some regions, such as the area surrounding Lagos -- the second most populous city in Africa -- showed a richness of species much higher in our simulations than is observed in reality. The simulation perhaps suggests that the area, with its tropical forest climate, might be very biodiverse were it not for the urban sprawl. More generally, Africa has seen widespread loss of native habitats due to deforestation and other destructive practices \citep{Steege2015, Gomes2019, Baccini2017, Perrings2015, Marengo2018}. Elevation also plays a major component in species richness patterns \citep{Barthlott2005} and, although changes in climate are already accounted for in these models, there are other potential drivers of elevational effects, including boundary conditions \citep{Colwell2004}. There are other additional complexities related to the modelling of different types of biomes, particularly species-rich rain forests, where canopy cover plays an important role in regulating population dynamics and access to limiting resources, such as light \citep{Meyer2019}. 

EcoSISTEM is an open-source tool, freely available for use and extension via GitHub \citep{Harris2021}, with several current applications beyond this continental biodiversity work, including simulating UK plant biodiversity under various climate and land-use change scenarios. Significant logistical challenges remain when modelling biodiversity at large scale, including parameterising this mechanistic, dynamic model for thousands or tens of thousands of species, extreme biases in available data on existing plant distributions, as well as computational logistics of running at such a large scale.

This work constitutes the first attempt at a truly species-level, dynamic, continental-scale model of plant biodiversity parameterised with real plant records and climate reconstructions, and tested on the African continent. Although much work remains, we already see that it can capture broad patterns in plant biodiversity for these areas, and it identifies areas that may have undergone diversity change throughout the course of the 20\textsuperscript{th} century. Though we are not yet able to say the models are predictive, future directions involve fitting of the model to smaller regions with more complete data, and we are already able to run the model through projections of future climate change. The largest comparable plant vegetation models are DGVMs, such as JULES, which work with up to 9 plant functional types \citep{Harper2016}. Though global in scope, a lack of species-level information hinders their capacity to describe and predict trends in plant biodiversity \citep{Mokany2016, Scheiter2013}. Other attempts to model communities of plants concentrate on individuals, such as forest gap models, or use statistical methods to model individual species separately and together, as in SDMs and joint SDMs, respectively. Scaling up to thousands of species across large areas comprises a considerable computational demand for all of these approaches \citep{Wilkinson2019, Shugart2020}.

In conclusion, here we present a simulator for plant biodiversity that includes demographic and competition processes and can be adapted for both size and type of habitat, as well as simulations of habitat and environmental change. Although the model is presented here with very incomplete information on the plant species simulated, it is nonetheless already parameterised on data from biological records and climate reconstructions. Using such data, leveraged from digital natural history collections, we show that changes in climate over the past century already have perceptible effects on the local plant biodiversity of Africa. This modelling approach also offers a route to exploit the value of natural history collections for novel information, including species traits and life history strategies.

\section*{Acknowledgements}
CH was supported by an Engineering and Physical Sciences Research Council (EPSRC) Studentship EP/M506539/1 ref. 1654080. RR was supported by the Biotechnology and Boiological Sciences Research Council (BBSRC) grant BB/P004202/1 and the Science and Technology Facilities Council (STFC) grant ST/V006126/1. CH, NB, CC and RR were supported by Natural Environment Research Council (NERC) grants NE/T004193/1 and NE/T010355/1. CH was also supported by the Scottish Government's Rural and Environment Science and Analytical Services Division (BioSS-3-LSM.)We would like to thank Anna Harper and Roderic Page for their insights into the paper. 
  
\section*{Author's contributions}

CH, RR, CC and NB conceived the ideas and designed the methodology. CH and RR developed the code base and CH led the writing of the manuscript. All authors contributed critically to the drafts and gave final approval for publication.

\section*{Data availability statement}
The source code for the analysis will be publicly available under an open source license on GitHub, and the release of the code associated with the paper will be provisioned with a DOI through Zenodo. The data used in the analysis are already publicly available (from ECMWF and GBIF).

\bibliographystyle{apacite}
\bibliography{bibs/Simulation}

\pagebreak

\section*{Supplementary Material}
\beginsupplement

\begin{table}[H]
\caption{Parameter details for case study 1: island and patch dynamics. Single values apply to all species, $s$, or grid cells, whereas others may vary within a given range.}
\footnotesize
\label{tab:climatedata}
\begin{center}
\ra{1.3}
\begin{tabulary}{\textwidth}{p{20em} p{12em}}
\toprule
 \textbf{Parameter name} & \textbf{Value} \\ 
\midrule
\textbf{Species} & \\
\midrule
Number of species  & 100 \\
Initial number of individuals & 100M \\
Average size & 1 m$^2$ \\
Sunlight requirement per individual, $\epsilon_{1,s}$ &  450000 kJ/m$^2$ \\
Water requirement per individual. $\epsilon_{2,s}$ & 192 nm/m$^2$ \\
Temperature preference, $\bar{T_s}$ & 293-303 K \\
Niche width, $\psi_s$ & 0.0001-5 K \\
Average dispersal distance, $\theta_s$ & 0.5-4 km \\
Birth rate, $b_s$ & 0.15/year \\
Death rate, $d_s$ & 0.15/year \\
Longevity, $\lambda$ & 1.0 \\
Survival, $\tau$ &  0.1 \\

\midrule
\textbf{Abiotic Environment} & \\
\midrule
Area & 100 km$^2$ \\
Grid cell width, $\eta$ & 1 km$^2$ \\
Total ecosystem sunlight availability, $K_1$ & 192 mm/km$^2$ \\
Total ecosystem water availability, $K_2$ & 4.5e$^{11}$ kJ/km$^2$ \\
Total ecosystem temperature, $T$ & 298 K \\
 \bottomrule
 
\end{tabulary}
\end{center}
\end{table}
\begin{table}[H]

\caption{Parameter details for case study 1: large scale dynamics. Single values apply to all species, $s$, or grid cells, whereas others may vary within a given range.}
\footnotesize
\label{tab:climatedata2}
\begin{center}
\ra{1.3}
\begin{tabulary}{\textwidth}{p{20em} p{12em}}
\toprule
 \textbf{Parameter name} & \textbf{Value} \\ 
\midrule
\textbf{Species} & \\
\midrule
Number of species  & 50,000 \\
Initial number of individuals & 3B \\
Average size & 1 m$^2$ \\
Sunlight requirement per individual, $\epsilon_{1,s}$ &  10 kJ/m$^2$ \\
Temperature preference, $\bar{T_s}$ & 274 K \\
Niche width, $\psi_s$ & 50 K \\
Average dispersal distance, $\theta_s$ & 15 km \\
Birth rate, $b_s$ & 0.6/year \\
Death rate, $d_s$ & 0.6/year \\
Longevity, $\lambda$ & 1.0 \\
Survival, $\tau$ &  0.1 \\

\midrule
\textbf{Abiotic Environment} & \\
\midrule
Area & 64M km$^2$ \\
Grid cell width, $\eta$ & 80 km$^2$ \\
Total ecosystem sunlight availability, $K_1$ & 1,000 kJ/km$^2$ \\
Total ecosystem temperature, $T$ & 274 K \\
 \bottomrule
 
\end{tabulary}
\end{center}
\end{table}

\begin{table}[H]
\caption{Parameter details for case study 2. Single values apply to all species, $s$, or grid cells, whereas others may vary within a given range.}
\footnotesize
\label{tab:climatedata3}
\begin{center}
\ra{1.3}
\begin{tabulary}{\textwidth}{p{20em} p{12em}}
\toprule
 \textbf{Parameter name} & \textbf{Value} \\ 
\midrule
\textbf{Species} & \\
\midrule
Initial number of species  & 39,158 \\
Initial number of individuals & 3B \\
Average size & 1 m$^2$ \\
Sunlight requirement per individual, $\epsilon_{1,s}$ &  From ECMWF data \\
Temperature preference, $\bar{T_s}$ & From ECMWF data \\
Niche width, $\psi_s$ & From ECMWF data \\
Average dispersal distance, $\theta_s$ & 0.6 - 2.4 km \\
Birth rate, $b_s$ & 0.15/year \\
Death rate, $d_s$ & 0.15/year \\
Longevity, $\lambda$ & 1.0 \\
Survival, $\tau$ &  0.01 \\

\midrule
\textbf{Abiotic Environment} & \\
\midrule
Area & 64M km$^2$ \\
Grid cell width, $\eta$ & 80 km$^2$ \\
Total ecosystem sunlight availability, $K_1$ & From ECMWF data \\
Total ecosystem temperature, $T$ & From ECMWF data \\
 \bottomrule
 
\end{tabulary}
\end{center}
\end{table}

\end{document}